\documentclass{aa}

\usepackage{amsmath}
\usepackage{graphicx}
\usepackage{graphics}
\usepackage{txfonts}
\begin{document}  


\title{A study of purely astrometric selection of extragalactic point sources with Gaia}

\author{
K.~E. Heintz\inst{1},
J.~P.~U. Fynbo\inst{1},
E. H{\o}g\inst{2}
}
\institute{
Dark Cosmology Centre, Niels Bohr Institute, Copenhagen University, Juliane Maries Vej 30, 2100 Copenhagen O, Denmark\\
\email{heintz@dark-cosmology.dk}
\and
Niels Bohr Institute, Copenhagen University, Juliane Maries Vej 30, 2100 Copenhagen O, Denmark
}

\date{Received 2015; accepted, 2015}

\abstract
{Selection of extragalactic point sources, e.g. QSOs, is often hampered by
significant selection effects causing existing samples to have rather complex
selection functions.} {We explore whether a purely astrometric selection of
extragalactic point sources, e.g. QSOs, is feasible with the ongoing Gaia
mission. The idea has been discussed in the context of Gaia, but it is
the first time quantified numbers have been given. Such a selection would be
interesting as it would be unbiased in terms of colours of the targets and
hence would allow selection also with colours in the stellar sequence.
}
{We have analyzed a 
total of 18 representative regions of the sky by using \textit{GUMS}, the
simulator prepared for ESAs Gaia mission, both in the range of $12\le G \le 20$
mag and $12\le G \le 18$ mag. For each region we determine the density of apparently
stationary stellar sources, i.e. sources for which Gaia cannot measure a significant 
proper motion. The density is contrasted with the density of extragalactic point sources,
e.g. QSOs, in order to establish in which celestial directions a pure astrometric 
selection is feasible.}
{When targeting regions at galactic latitude $|b| \ge 30^\mathrm{o}$ the
ratio of QSOs to apparently stationary stars is above 50\% and when observing
towards the poles the fraction of QSOs goes up to about $\sim80$\%. We show that the proper motions from the proposed Gaia successor mission in about 20 years would dramatically improve these results at all latitudes.
Detection of QSOs solely from zero proper motion,
unbiased by any assumptions on spectra, might lead to the discovery of new
types of QSOs or new classes of extragalactic point sources. 
}
{}

\keywords{quasars: general -- Astrometry -- Proper motions}

\maketitle

\section{Introduction}     \label{sec:introduction}

Since their discovery in the early 1960s \citep{Schmidt63} numerous surveys for
Quasi Stellar Objects (QSOs) have been carried out and the number of known QSOs
now count hundreds of thousands.
To have a complete understanding of super-massive black hole formation and 
evolution it is desirable to have QSO samples selected in several different ways 
in order to be less constrained by selection effects.

The incompleteness of QSO samples based on selection by
optical photometry has been studied intensively for many years 
\citep[see, e.g.,][for a color selection of QSOs from the SDSS survey]{Richards2004} 
and it is now well established that such samples miss a substantial number of in
particular dust-reddened QSOs \citep[see, e.g.,][for a recent study]{Krawczyk}.
 
In order to examine the feasibility
of this approach we first need to determine the contamination by the stars in
our galaxy, by analyzing how many stars will be selected by the zero proper
motion criterion and towards which galactic coordinates the problem of stellar
contamination will be most severe.\\\\ 
To describe this we have structured the
paper as follows. We start with a brief description of the Gaia mission and
show the expected errors on proper motion from this mission and the usefulness 
of the soon to be
obtained highly accurate astrometric measurements in Sect.~\ref{sec:data}. To get a hint
on these future data we have in this paper used extracts from a catalog of 1.6
billion stars generated for the Gaia mission, the so-called Gaia Universe model
snapshot (GUMS) which also will be shortly described in Sect.~\ref{sec:data}. 
In Sect.~\ref{sec:results} and \ref{sec:conclusions} we will discuss the results of this analysis and thereby conclude
on the feasibility of this new approach. 

In the analysis we not only consider
what we can accomplish with the Gaia mission alone, but also with proper motions having 10 times smaller errors that from Gaia. This accuracy can be obtained by a combination of positions from Gaia and from another all-sky astrometric space mission in 20 years with similar errors on positions as Gaia; in fact this would be the only feasible way to obtain such accuracy. We will in the following refer to
such a second mission as a "Gaia successor".

\section{Simulated Gaia Data}    \label{sec:data}
The Gaia mission, launched by ESA
in December 2013, is a very powerful astrometric mission that is scheduled to make a
three-dimensional map of our galaxy, and provide an unprecedented measurement
of positions, proper motions and parallaxes to more than one billion stellar systems to a
limit in the G-band of 20 mag during its 5-6 years mission life time \citep{deBruijne2012}.

\subsection{GUMS}
Gaia is expected to transmit close to 150 terabytes of raw data, therefore
preparation of acquiring this amount of data is essential \citep{Luri2014}.
Hence, the Gaia Data Processing and Analysis Consortium (DPAC) has produced a
set of simulators, including the Gaia Object Generator (GOG) which provides
simulations of number counts and lists of observable objects and is designed to
simulate catalog data \citep[see, e.g.,][]{Robin2012}. 

A basic component of the
Gaia simulator is its Universe Model (UM) from 
\citet{Robin2003,Robin2004} and this
model is capable of simulating almost every object down to Gaia's limiting
magnitude of $G=20$ mag, both for galactic and extra-galactic objects (Luri et
al. 2014). The Gaia simulator combined with the universe model is then supposed
to show a snapshot of the potentially observable objects by Gaia, thereby
called the Gaia Universe Model Snapshot (GUMS). See especially Sect.~3 in
Robin et al. (2012) for the full description of the stellar content. 

We have
used the extracts from this catalog of 1.6 billion stars which have been
generated by GUMS and can be obtained via the vizier website 
\footnote{{\tt http://vizier.u-strasbg.fr/viz-bin/VizieR-3}}. This has enabled us to
derive precise numbers for the expected stellar contamination and the probability of
separating QSOs and other extragalactic point sources from galactic sources.

\subsection{Expected Gaia errors}
The standard errors for a five year mission as expected before the launch of Gaia have been assumed here. These errors and the signal-to-noise ratio, $S/N$, are calculated as a function of the apparent magnitude G, which is approximately equal to the visual magnitude V. The following formulae are taken from \citet{deBruijne2012}\footnote{{\tt functions listed at: http://www.cosmos.esa.int/web\\
/gaia/science-performance}}:
\begin{align}
\begin{split}
z &= 10^{(0.4\cdot(G-15))},~\mathrm{for~G>12~mag}\\
\sigma_{\pi}~[\mu\mathrm{as}] &= \left({9.3+658.1\cdot z + 4.568\cdot z^2}\right)^{1/2}\\
&\times~\left[0.986 + (1-0.986)\cdot(V-I_C)\right],\\
\sigma_{\mu}~[\mu\mathrm{as}~\mathrm{yr}^{-1}] &= 0.526\cdot\sigma_{\pi}, ~~~ S/N = \frac{\mu}{\sigma_{\mu}}, 
\end{split}
\end{align}
where 
$G$ is the \textit{G}-band magnitude,
$\sigma_{\pi}$ denotes the error on the parallax measurement,
$\sigma_{\mu}$ is the error on the proper motion,
and for an unreddened G2V star we can set $(V-I_C)=0.75$ \citep{deBruijne2012}. 
For $G=20$ mag this yields a lower limit in $\sigma_{\mu}$ of about 0.3 $mas/yr$ and for $G=18$ mag it is close to 0.1 $mas/yr$.

\section{Results}    \label{sec:results}
It is now possible to determine the expected contamination from stars with
proper motions below the detection limit of Gaia.

\begin{figure}[ht]
    \centering
    \includegraphics[scale=0.5]{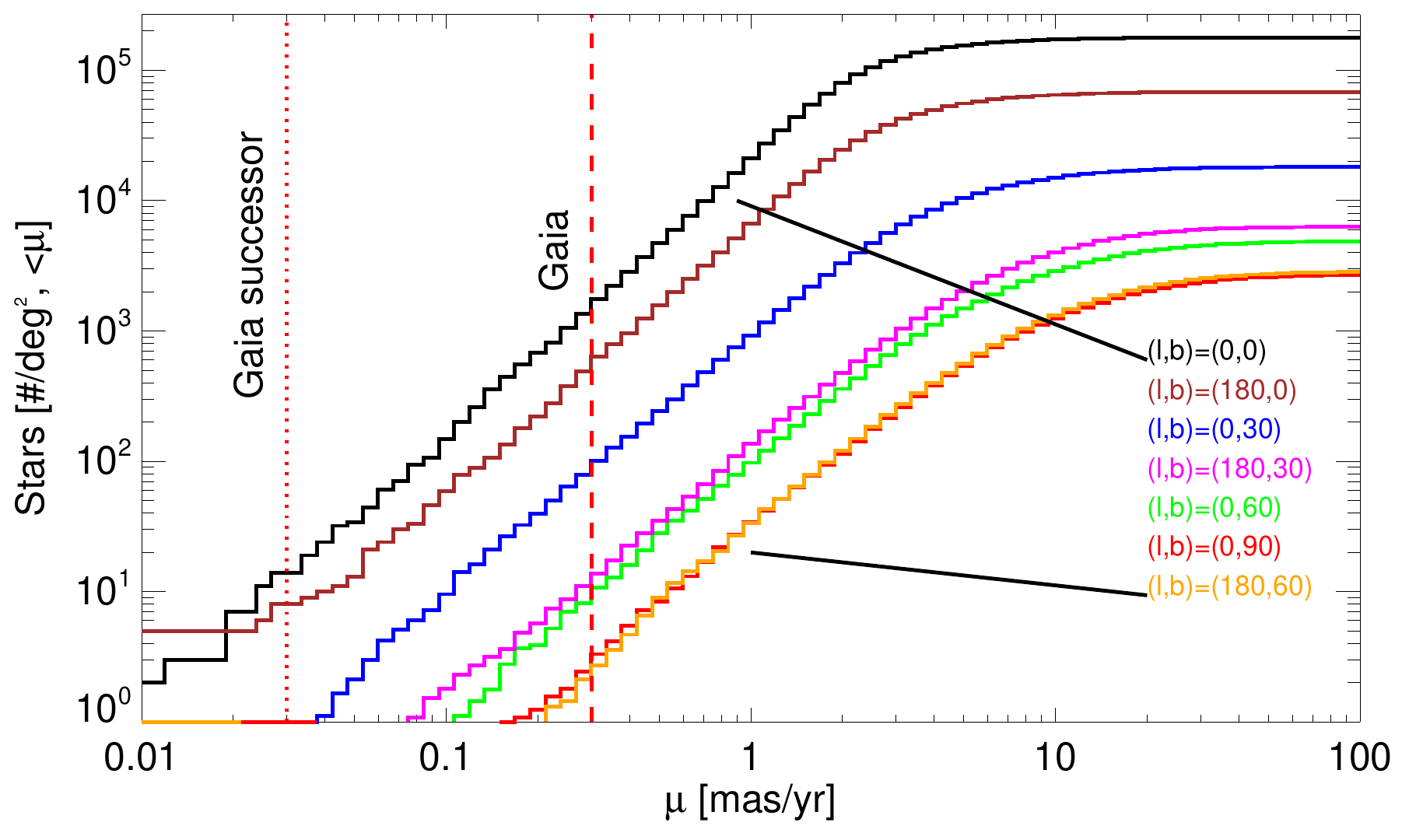}
    \caption{This figure shows the number of stars with a magnitude range in the G-band of $12 \le G \le 20$ per square degree versus the cumulative distribution of each of their proper motion as seen on the sky, listen in 7 directions as given by the galactic coordinates (\textit{l,b}). The values (\textit{l,b}) are listed at the right in the same sequence as the curves are seen. The red dashed line at $\mu = 0.3~mas/yr$ represents the estimated error on the proper motions from Gaia at $G=20$ mag. The red dotted line at $\mu = 0.03~mas/yr$ represents the estimated error on the proper motions derived from the positions obtained from Gaia and the proposed Gaia successor mission, also at $G=20$ mag.}
    \label{fig:lat20mu}
\end{figure}

In Fig.~\ref{fig:lat20mu} we have plotted the expected number of $G \le 20$ stars per square degree with proper motion less than $\mu$
as a function of $\mu$ in 7 directions, of a total of 18 directions, as listed in Tables \ref{tab:long20} to \ref{tab:lat18} of Sect.~\ref{sec:results}. Note, that for the plots of varying galactic
latitude, \textit{b}, we extracted a larger area of the sky of 3$\times$3 or 5$\times$5 deg$^2$, and
afterwards divided the total number of stars with 9 or 25, respectively, in
order to normalize the y-axis since the 1$\times$1 deg$^2$ area away from the galactic
plane has a low density of stars.

\begin{figure}[ht]
    \centering
    \includegraphics[scale=0.51]{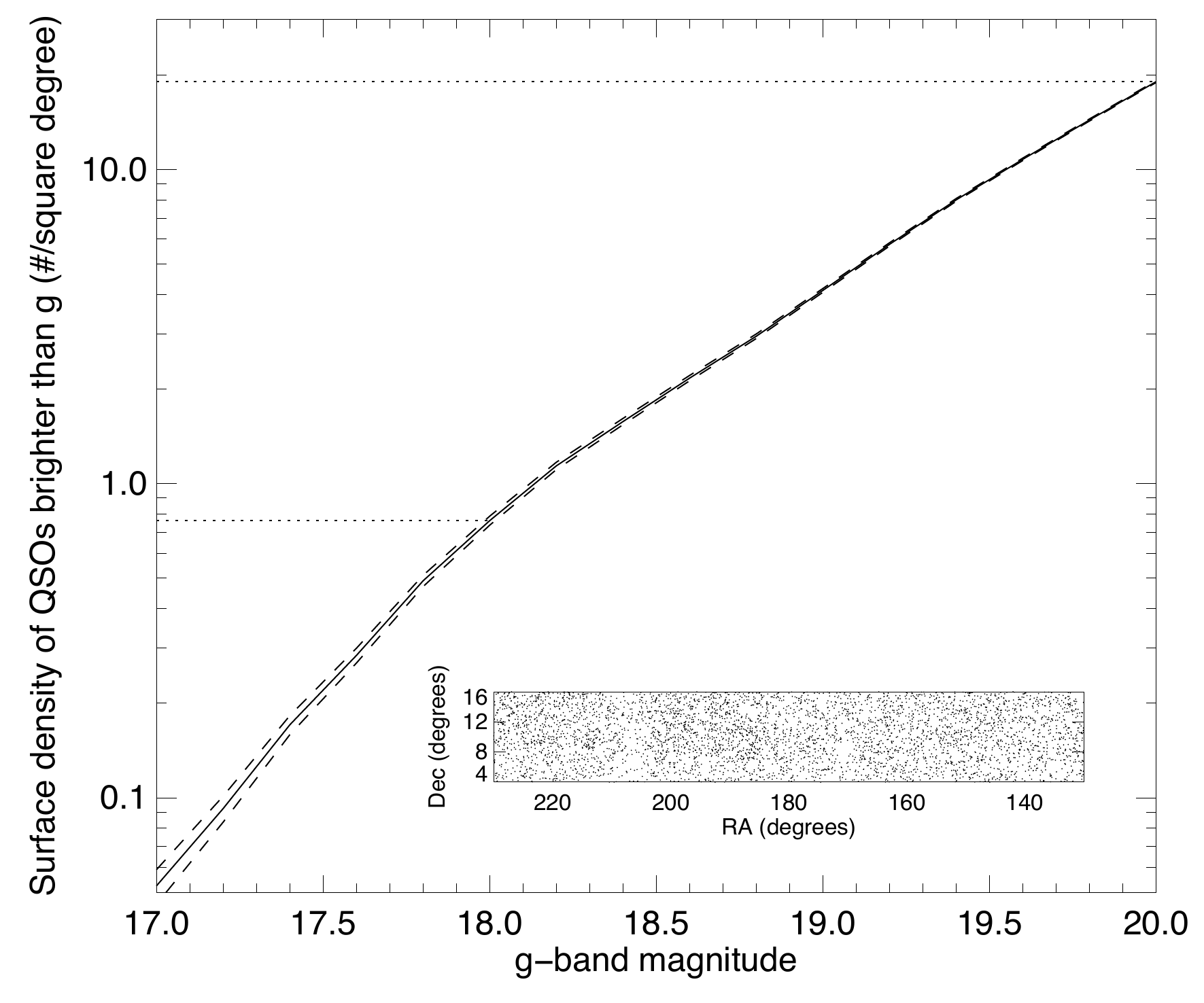}
    \caption{Plot of surface densities of QSOs at given magnitudes in the
    g-band from the BOSS catalog. The insert shows the celestial region we have
    used to generate the plot, i.e. a region well away from the Galactic disk
    where the distribution is uniform. The two dotted lines show the QSO surface
    density for one square degree, which is found to be $19$ and $0.76$ when
    observing at a limited range of $g=20$ and $g=18$ mag, respectively. In the present context we consider the bands g and G to be equivalent although the effective wavelengths differ by about 100 nm, but they vary greatly with spectral type because both bands are very wide.}
    \label{fig:BOSSQSO}
\end{figure}

\begin{figure}[ht]
    \includegraphics[scale=0.51]{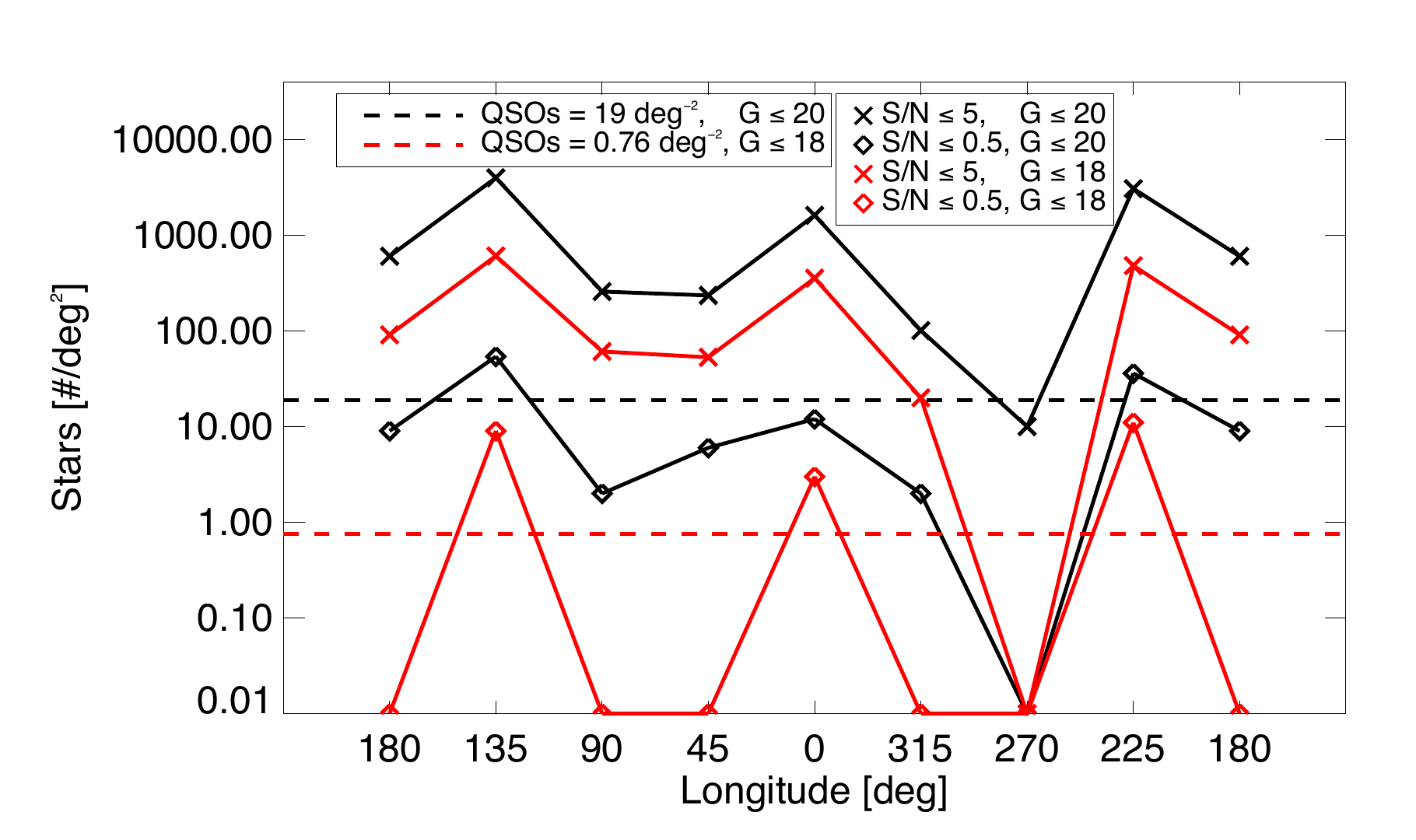}
    \includegraphics[scale=0.51]{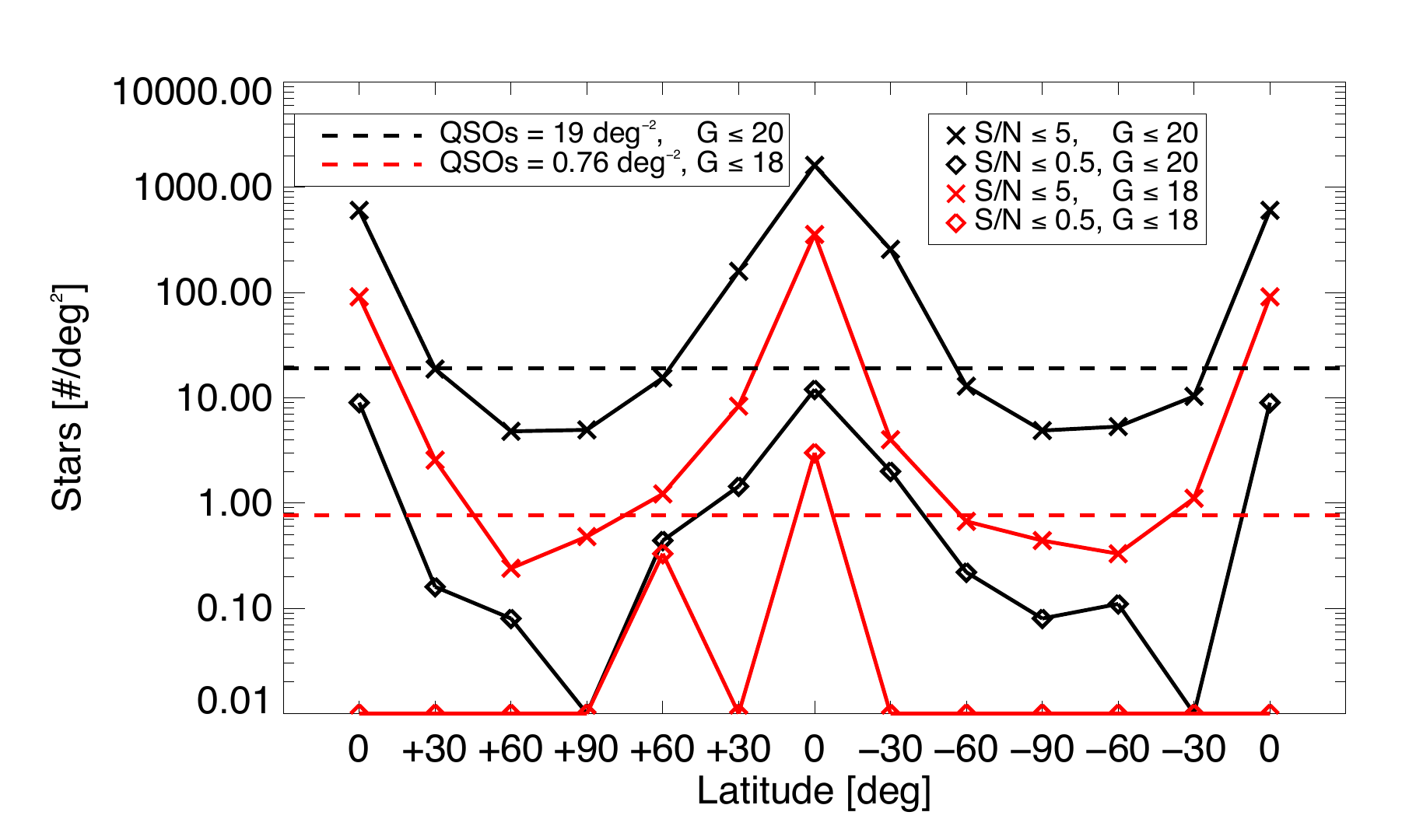}
    \caption{{\it Top panel}: the results of the analyzed stellar proper
    motions, when observing at $b=0$ and varying galactic longitudes, with a $S/N \le 5$ for
    Gaia (marked with crosses) and a $S/N \le 0.5$ for a combination of Gaia and a proposed Gaia
    successor mission (marked with diamonds), both for $G \le 20$ mag and $G \le 18$ mag (shown with black
    and red lines, respectively). {\it Bottom Panel}:
    the results of the analyzed stellar proper
    motions, when observing at $b=0$ and varying galactic latitudes, with a $S/N \le 5$ for
    Gaia and a $S/N \le 0.5$ for a combination of Gaia and a proposed Gaia
    successor mission, both for $G \le 20$ mag and $G \le 18$ mag (with same 
    use of crosses, diamonds, black and red lines as in the top panel). Black and red dashed lines represent the number of QSOs at $G=20$ and $G=18$ mag, respectively, from the BOSS catalog for comparison.}
    \label{fig:longresbr}
\end{figure}

\begin{table*}[!htbp]
\centering
\begin{minipage}{0.7\textwidth}
\caption{Results of data when only varying $l$, in the galactic plane for $G \le 20$ mag.}
\scalebox{0.9}{
    \begin{tabular}{lcccccccc}
    \noalign{\smallskip} \hline \hline \noalign{\smallskip}
        \emph{Longitude [deg]} & \emph{180} & \emph{135} & \emph{90} & \emph{45} & \emph{0} & \emph{315} & \emph{270} & \emph{225} \\
        \hline
        $N_{tot}$ & 68375 & 86061 & 210314 & 200141 & 178152 & 196128 & 26844 & 95276 \\
        $N_{\mu,0.3}$ & 2254 & 12631 & 1007 & 991 & 6917 & 358 & 61 & 9712 \\
        $N_{SN,5}$ & 604 & 3991 & 258 & 238 & 1616 & 101 & 10 & 3072 \\
        $N_{SN,0.5}$ & 9 & 54 & 2 & 6 & 12 & 2 & 0 & 36 \\
        $N_{QSO}/(N_{SN,5}+N_{QSO})$ & 0.030 & 0.005 & 0.069 & 0.074 & 0.012 & 0.158 & 0.655 & 0.006 \\
        $N_{QSO}/(N_{SN,0.5}+N_{QSO})$ & 0.679 & 0.260 & 0.905 & 0.760 & 0.613 & 0.905 & 1 & 0.345 \\
        \noalign{\smallskip} \hline \noalign{\smallskip}
    \end{tabular}}
    \label{tab:long20}
    \end{minipage}
    \end{table*}

\begin{table*}[!htbp]
\centering
\begin{minipage}{0.9\textwidth}
\caption{Results of data when only varying $b$, observing away from the galactic plane for $G \le 20$ mag, beginning at anticenter.}
\scalebox{0.9}{
    \begin{tabular}{lcccccccccccc}
    \noalign{\smallskip} \hline \hline \noalign{\smallskip}
        \emph{Latitude [deg]} & \emph{(180,0)} & \emph{+30} & \emph{+60} & \emph{+90} & \emph{+60} & \emph{+30} & \emph{(0,0)} & \emph{-30} & \emph{-60} & \emph{-90} & \emph{-60} & \emph{-30}  \\
        \hline
        $N_{tot}$ & 68375  & 6323 &  2851 & 2720  & 4889 & 18145 & 178152 & 25244 & 5174 & 2852 & 2953 & 5511 \\
        $N_{\mu,0.3}$ & 2254 & 48.7 & 13.1 & 12.0 & 38.9 & 345.3 & 6917 & 544 & 32.11 & 11.52 & 12.11 & 36 \\
        $N_{SN,5}$ & 604 & 18.76 & 4.80 & 4.96 & 15.44 & 159 & 1616 & 257 & 12.89 & 4.88 & 5.33 & 10.33 \\
        $N_{SN,0.5}$ & 9 & 0.16 & 0.08 & 0 & 0.44 & 1.44 & 12 & 2 & 0.22 & 0.08 & 0.11 & 0 \\
        $N_{QSO}/(N_{SN,5}+N_{QSO})$ & 0.030 & 0.503 & 0.798 & 0.793 & 0.552 & 0.107 & 0.012 & 0.069 & 0.596 & 0.796 & 0.781 & 0.648 \\
        $N_{QSO}/(N_{SN,0.5}+N_{QSO})$ & 0.679 & 0.992 & 0.996 & 1 & 0.977 & 0.930 & 0.613 & 0.905 & 0.989 & 0.996 & 0.994 & 1 \\
        \noalign{\smallskip} \hline \noalign{\smallskip}
    \end{tabular}}
    \label{tab:lat20}
    \end{minipage}
    \end{table*}

\begin{table*}[!htbp]
\centering
\begin{minipage}{0.7\textwidth}
\caption{Results of data when only varying $l$, in the galactic plane for $G \le 18$ mag.}
\scalebox{0.9}{
    \begin{tabular}{lcccccccc}
    \noalign{\smallskip} \hline \hline \noalign{\smallskip}
        \emph{Longitude [deg]} & \emph{180} & \emph{135} & \emph{90} & \emph{45} & \emph{0} & \emph{315} & \emph{270} & \emph{225} \\
        \hline
        $N_{tot}$ & 40156 & 48071 & 110018 & 115174 & 107542 & 113205 & 16761 & 53235 \\
        $N_{\mu,0.1}$ & 125 & 771 & 77 & 65 & 426 & 27 & 4 & 598 \\
        $N_{SN,5}$ & 91 & 608 & 61 & 53 & 356 & 20 & 1 & 481 \\
        $N_{SN,0.5}$ & 0 & 9 & 0 & 0 & 3 & 0 & 0 & 11 \\
        $N_{QSO}/(N_{SN,5}+N_{QSO})$ & 0.008 & 0.001 & 0.012 & 0.014 & 0.002 & 0.037 & 0.432 & 0.002 \\
        $N_{QSO}/(N_{SN,0.5}+N_{QSO})$ & 1 & 0.078 & 1 & 1 & 0.202 & 1 & 1 & 0.065 \\
        \noalign{\smallskip} \hline \noalign{\smallskip}
    \end{tabular}}
    \label{tab:long18}
    \end{minipage}
    \end{table*}

\begin{table*}[!htbp]
\centering
\begin{minipage}{0.9\textwidth}
\caption{Results of data when only varying $b$, observing away from the galactic plane for $G \le 18$ mag, beginning at anticenter.}
\scalebox{0.9}{
    \begin{tabular}{lcccccccccccc}
    \noalign{\smallskip} \hline \hline \noalign{\smallskip}
        \emph{Latitude [deg]} & \emph{(180,0)} & \emph{+30} & \emph{+60} & \emph{+90} & \emph{+60} & \emph{+30} & \emph{(0,0)} & \emph{-30} & \emph{-60} & \emph{-90} & \emph{-60} & \emph{-30}  \\
        \hline
        $N_{tot}$ & 40156 & 3761 & 1673 & 1542 & 2564 & 7938 & 107542 & 10828 & 2731 & 1629 & 1744 & 3360 \\
        $N_{\mu,0.1}$ & 125 & 2.76 & 0.44 & 0.52 & 1.22 & 10.33 & 426 & 12 & 1 & 0.20 & 0.33 & 1.78 \\
        $N_{SN,5}$ & 91 & 2.56 & 0.24 & 0.48 & 1.22 & 8.33 & 356 & 4 & 0.67 & 0.44 & 0.33 & 1.11 \\
        $N_{SN,0.5}$ & 0 & 0 & 0 & 0 & 0.33 & 0 & 3 & 0 & 0 & 0 & 0 & 0 \\
        $N_{QSO}/(N_{SN,5}+N_{QSO})$ & 0.008 & 0.229 & 0.760 & 0.613 & 0.384 & 0.084 & 0.002 & 0.160 & 0.531 & 0.633 & 0.697 & 0.406 \\
        $N_{QSO}/(N_{SN,0.5}+N_{QSO})$ & 1 & 1 & 1 & 1 & 0.697 & 1 & 0.202 & 1 & 1 & 1 & 1 & 1 \\
        \noalign{\smallskip} \hline \noalign{\smallskip}
    \end{tabular}} 
    \label{tab:lat18}
    \end{minipage}
    \end{table*}

To be conservative, we consider stars with proper motions measured with $S/N \le 5$ as
possible contaminants. We also calculate the number of sources with $S/N \le 0.5$,
which is the equivalent of $S/N \le 5$ for a data set based on Gaia plus a successor 
mission similar to Gaia operating about 20 years from now.

Such a Gaia successor mission has been proposed in May 2013 by \citet{Hoeg2014a} as being crucial for the astrometric foundation of astrophysics and the proposal has been further refined in \citet{Hoeg2014b}.

We show the results found when analyzing how many of the total
number of stars in each direction in one square degree have $S/N \le 5$ and
$S/N \le 0.5$, both for $G \le 20$ mag and $G \le 18$ mag.  In Fig.~\ref{fig:longresbr} 
this is shown graphically, whereas
the listed values are shown in Tables~\ref{tab:long20}, \ref{tab:lat20},
\ref{tab:long18} and  \ref{tab:lat18}. The points for $G=20~\mathrm{mag}$, shown with black points in Fig.~\ref{fig:longresbr}, 
show that the number of contaminating stars is reduced by a factor 100 with the proposed Gaia successor mission, precisely what was expected with 10 times smaller errors of proper motions in the two celestial coordinates.

The surface density of known QSOs at each
limiting magnitude from the BOSS catalog \citep[see, e.g.,][for the full catalog 
description]{Paris2012} is plotted in Fig.~\ref{fig:BOSSQSO}. Note that here we have
plotted as a function of $g$-band magnitude as we do not have $G$-band magnitudes
for the BOSS QSOs.
The density of BOSS QSOs is shown as the two dotted lines (assuming $g \approx G$) in 
Fig.~\ref{fig:longresbr} to indicate the relative frequency of known
QSOs and galactic stars for each pointing direction.

It can be concluded from Fig.~\ref{fig:longresbr} that when observing with Gaia at $G \le 20$ mag, 
the probability that a point source with proper motion detected at less than 5$\sigma$ is a QSO drops rapidly when 
observing below $|b|= 30^{\mathrm{o}}$. We have the highest probability to select a QSO with Gaia at 
$S/N \le 5$ for $G \le 20$ at $(l,b)=(180,60)$ where, according to Table \ref{tab:lat20}:
\begin{equation}
\frac{N_{QSO}}{(N_{QSO}+N_{SN,5})}=0.798=79.8\%.
\end{equation}
The probability is just above 50\% at $|b| = 30^{\mathrm{o}}$ which means that this way of selecting 
extragalactic point sources clearly is feasible. Also shown in Fig.~\ref{fig:longresbr} is that
with a Gaia successor mission 20 years after the current Gaia mission the contamination from apparently stationary
stars essentially would be eliminated.

When restricting the analysis to a brighter limiting magnitude of
$G \le 18$ instead, the contamination from apparently stationary stars
is slightly higher. The first line of Tables~\ref{tab:long20} to
\ref{tab:lat18} denotes the total number of stars per square degree in the
listed direction. The next three lines show the number of stars with values
lower than $\mu \le 0.3~mas/yr$, $S/N \le 5$ and $S/N \le 0.5$, respectively.
The final two lines shows the QSO fraction, when observing point sources in the
limit of $S/N = 5$ and $S/N = 0.5$, respectively. The numbers in last line of 
Table~\ref{tab:long20} demonstrate that even in the galactic plane QSOs can be extracted with high probability at the longitudes 90, 315 , and 270 degrees, but these numbers do not take interstellar absorption into account.

\section{Discussion and Conclusions} \label{sec:conclusions}

We have used the object simulator prepared for the Gaia mission, GUMS, to
determine the expected contamination from galactic stars in a search for
extragalactic point sources based on astrometric measurements from Gaia.

By analysing the GUMS we determine the frequency of selecting a QSO
based on the density inferred from BOSS against the background of 
apparently stationary stars. This provides a conservative estimate of the
relative number of QSOs and apparently stationary stars as the BOSS 
survey does not contain all QSOs down to the flux limit of the 
survey \cite[examples of missed QSOs can be found in][]{Fynbo13,Krogager14}.

At $|b|>30^{\mathrm{o}}$ the ratio
of QSOs relative to stars is above 50\% when observing down to 
$G \le 20$ mag.
When decreasing the magnitude
range to $G \le 18$ mag a lower number of stellar contamination is
obtained but the relative QSO surface density in that magnitude range decreases
as well, yielding a sightly higher level of stellar contamination. When
observing below $|b|= 30^{\mathrm{o}}$, close to the galactic plane, the
contamination increases rapidly.

With a Gaia successor mission, i.e. a mission similar to Gaia launched in
about 20 years from now, the standard error on
the measurement of the proper motions would decrease by about a factor of
ten, resulting in 100 times improvement in the ability to 
select extragalactic point sources against the background of apparently
stationary stars.

Since the method we analyse here is unbiased in terms of colour 
this method has the potential to discover new,
exotic types of QSOs and even in principle new classes of extragalactic point sources. 
Also, this method would allow the construction of an unbiased QSO sample limited only
in its flux limit. This would be very valuable for adressing several issues,
e.g. 
the true metallicity distribution of foreground Damped Lyman-$\alpha$
Absorber galaxies \cite[e.g.,][]{Fall93} or the redshift distribution of
Broad Absorption Line QSOs \citep[][]{Saturni15}. 

We also note that the astrometric information from Gaia will be 
very useful to remove contaminating stellar sources from other more
targetted searches for red QSOs. Often such contaminating sources
are M-dwarfs, which due to their small distances should have significant
proper motions.

We have assumed that QSOs are point sources, but we are aware that this is not always the case. Variability of a QSO with changing centres could introduce observed proper motions. We are aware that proper motions of quasars may be significant. In the catalog by \citet{Titov2011}  with 555 radio sources the motions are typically 0.1 mas/yr and some even 1 mas/yr, but the 40 most frequently observed sources shown in Fig.2 of the v4-version on arXiv of the paper all have much smaller motions of less than 0.025 mas/yr. One can argue that optical QSOs should be more stable than radio sources, more compact etc. But we don't know for sure until Gaia data become available.
%

\begin{acknowledgements}
The data from GUMS, the Gaia Universe Model Snapshot, were aquired with kind help from Carine Babusiaux and they were used for simulation of observations as expected from Gaia.
We thank Coryn Bailer-Jones, Sergei Klioner and Palle M\o ller for helpful comments.
The Dark Cosmology Centre is funded by the DNRF. The
research leading to these results has received funding from the European
Research Council under the European Union's Seventh Framework Program
(FP7/2007-2013)/ERC Grant agreement no.  EGGS-278202.
\end{acknowledgements}

\bibliographystyle{aa}
\bibliography{ref}

\end{document}